\documentclass[12pt]{article}

\usepackage{float}
\usepackage[makeroom]{cancel}
\usepackage{color}
\usepackage{graphicx}
\usepackage{amsmath}
\usepackage{amssymb}
\usepackage{xspace}
\usepackage[small]{subfigure}
\usepackage[numbers,compress]{natbib}
\usepackage[hyperfootnotes=false]{hyperref}
\usepackage{tcolorbox}

\usepackage{framed}
\usepackage{physics}
\usepackage{tensor}

\newlength{\fighskip} \fighskip=2pt
\newlength{\figvskip} \figvskip=3pt

\newcommand*{\figbox}[2]{{
  \def\figscale{#1}
  \def\arraystretch{0.8}
  \arraycolsep=0pt
  \begin{array}{c}
    \vbox{\vskip\figscale\figvskip
      \hbox{\hskip\figscale\fighskip
        \includegraphics[scale=\figscale]{#2}}}
  \end{array}}}

\usepackage{mciteplus} 
\usepackage{dcolumn}
\usepackage{bm}
\usepackage{verbatim}
\usepackage{amscd}
\usepackage{amsfonts}
\usepackage{setspace}
\usepackage{amsthm}
\usepackage{enumerate}
\usepackage{mathtools}

\newtheorem{theorem}{Theorem}
\theoremstyle{plain}

\theoremstyle{plain}

\DeclareMathOperator{\ED}{ED}

\newcommand{\prob}[1]{\mathbb{P}\left(#1\right)}

\newcommand{\defeq}{\stackrel{\text{def}}{=}}

\usepackage{authblk}

\title{
\vspace{-80pt}
\hfill
{\normalsize RUP-25-23}\\
\vspace{40pt}
\bf 
Baby universe as logical qubits:
\\
information recovery in random encoding
}
\author[1]{Takato Mori\thanks{takato.mori@yukawa.kyoto-u.ac.jp}}
\author[2]{Beni Yoshida\thanks{byoshida@perimeterinstitute.ca}}
\affil[1]{\em \small Department of Physics, Rikkyo University, \protect\\
3-34-1 Nishi-Ikebukuro, Toshima-ku, Tokyo 171-8501, Japan}
\affil[2]{\em \small Perimeter Institute for Theoretical Physics, Waterloo, Ontario N2L 3W8, Canada}
\date{}

\begin{document}
\maketitle

\begin{abstract}
We revisit whether a semiclassical closed baby universe in AdS/CFT necessarily possess a trivial one-dimensional Hilbert space or may instead carry a large entropy. 
Recent results on Haar random encoding suggest a breakdown of complementary recovery, in which no logical operators can be reconstructed from individual bipartite subsystems.  
Motivated by this, we propose an interpretation where a baby universe emerges as logical degrees of freedom that cannot be accessed from either boundary alone, assuming pseudorandom dynamics in holographic CFT correlators. 
We then analyze two conceptual puzzles: an apparent cloning of baby-universe microstates and its eventual fate at the singularity. 
Both puzzles are avoided because no single boundary observer can access the baby-universe degrees of freedom, be it classical or quantum, reflecting an emergent form of complementarity due to the structure of random encoding. 
In this interpretation, observers arise naturally: the same heavy operator that prepares the baby-universe geometry also serves as observer-like degrees of freedom that define an observer-dependent baby-universe microstate. 
\end{abstract}

\newpage

\section{Introduction}

Does a closed baby universe possess a trivial one-dimensional Hilbert space with nothing to explore, or can it accommodate a large entropy and host rich internal physics?
Recent developments in gravitational path integrals suggest the existence of semiclassical gravitational saddles describing joint states of two asymptotically AdS regions entangled with a semiclassical
baby universe~\cite{Antonini:2023hdh}. 
However, when non-perturbative wormhole effects are included, the same path integrals indicate an extremely small, possibly one-dimensional, Hilbert space associated with the baby universe~\cite{Abdalla:2025gzn}.  
While a semiclassical baby universe might in principle be captured by analyzing the boundary CFTs beyond the naive large-$N$ limit, it remains obscure how, or even where, the baby-universe Hilbert space is encoded within the boundary Hilbert spaces. 

Quantum error correction provides a framework to analyze how quantum information can be encoded non-locally in many-body entangled systems. 
This perspective has proven invaluable in quantum gravity, where information is scrambled and delocalized by chaotic dynamics~\cite{Hayden:2007cs, Hosur:2015ylk}. 
Although a complete resolution of the baby-universe puzzle ultimately requires a full theory of quantum gravity, the language of quantum error correction offers a precise yet intuitive framework to clarify underlying assumptions and illuminate the physical mechanisms behind it.

In this paper, we construct a simple toy model of quantum error-correcting codes that capture essential features of  how a baby universe may emerge in two-sided CFTs. 
A key property of the semiclassical baby universe from~\cite{Antonini:2023hdh} is that it resides \emph{outside the entanglement wedges} of both CFTs on the boundary. 
At first sight, this seems to support the idea that the baby universe is absent, or that its Hilbert space is trivial. 
However, it also leaves open the possibility that the baby universe may be encoded as logical qubits in a quantum error-correcting code where its codeword states remain inaccessible to either boundary individually, but recoverable through non-local joint operations involving both boundaries. 

Concretely, we consider a Haar random encoding with an isometry
\begin{align}
V : C\rightarrow AB
\end{align}
where $A$ and $B$ denote the two boundary CFTs, and $C$ represents bulk degrees of freedom to be reconstructed from them. 
This random quantum error-correcting code exhibits the characteristic behavior expected of a baby universe. 
When the Hilbert space dimensions satisfy 
\begin{align}
d_A < d_B d_C ,\qquad d_B < d_A d_C,
\end{align}
no quantum information about $C$ can be retrieved from either $A$ or $B$ alone, as neither subsystem supports any non-trivial unitary logical operators of the code. 
This property follows from recent results on the absence of bipartite entanglement in typical tripartite Haar random states~\cite{Li:2025nxv}. 
From the viewpoint of each boundary, the Hilbert space $C$ appears absent, yet it still exists as logical qubits accessible only through joint, non-local operations acting on $A$ and $B$ together. 

This property of a random encoding has a somewhat surprising implication: the \emph{breakdown of complementary recovery}. 
Roughly speaking, complementary recovery refers to a special class of quantum error correction in which the input Hilbert space $C$ can be bipartitioned as $C\simeq C_A\otimes C_B$ such that $C_A$ and $C_B$ are reconstructable from $A$ and $B$ respectively~\cite{Harlow:2016vwg}. 
Complementary recovery holds in most known examples of AdS/CFT and also in stabilizer codes when formulated through operator-algebraic frameworks. 
In contrast, a Haar-random encoding exhibits a \emph{complete} breakdown of complementary recovery where no logical operator can be reconstructed on any bipartite subsystem.~\footnote{
The HaPPY code is constructed with stabilizer tensors, and thus satisfies complementary recovery~\cite{Pastawski:2015qua}. On the contrary, a random tensor network code~\cite{Hayden:2016cfa} is constructed with Haar random tensors, and thus violate complementary recovery in certain cases~\cite{Mori:2024gwe}.
}

Equipped with this toy model, we revisit the semiclassical baby universe setup by Antonini, Sasieta, and Swingle (AS$^2$) to examine how a pseudorandom encoding may arise~\cite{Antonini:2023hdh}. 
In the AS$^2$ construction, a shell of heavy operators $O$ is inserted and evolved in Euclidean time by $e^{-\frac{\beta H}{2}}$. 
Here we generalize to a family of randomly-chosen shells of heavy operators $\mathcal{M}_{\mathrm{heavy}} = \{O^{(i)}\}$, and study the corresponding ensemble of boundary states. 
Each insertion constructs a two-sided CFT wavefunction expanded in the low-energy basis below the Hawking-Page temperature as
\begin{align}
|\psi_{AB}^{(i)} \rangle = \frac{1}{\sqrt{Z^{(i)}}} \sum_{m,n} \tilde{O}^{(i)}_{mn}|m\rangle_{A} \otimes |n\rangle_{B}
\end{align}
where $\tilde{O}^{(i)}_{mn}$ are matrix elements of the Euclidean-evolved heavy operator. 
We suggest that this tensor $\tilde{O}^{(i)}_{m,n}$ exhibits pseudorandom features akin to those of a tripartite Haar random state. 
This expectation rests on an assumption that holographic CFT correlation functions involving mixtures of heavy and light operators display pseudorandom statistics due to underlying chaotic dynamics of holographic CFTs~\cite{Lashkari:2016vgj, Collier:2019weq, Belin:2020hea}.

To make this structure explicit, we introduce a fictitious third boundary system $C$ entangled with the label $i$ of the heavy operator:
\begin{align}
|\Psi_{ABC} \rangle = \frac{1}{\sqrt{Z}} \sum_{i,m,n} \tilde{O}^{(i)}_{mn}|m\rangle_{A} \otimes |n\rangle_{B} \otimes |i\rangle_C.
\end{align}
This construction suggests that the baby-universe Hilbert space can be regarded as the code subspace of a random isometry, $C \rightarrow AB$. 
Namely, projecting onto a particular state of $C$ prepares a codeword state on $AB$ which may be regarded as a specific baby-universe microstate.
If one’s access is limited to local operators on $A$ or $B$, no information about $C$ can be recovered.
Then the boundary observer perceives a mixed state,
\begin{align}
\rho_{AB} = \sum_i |\psi_{AB}^{(i)}\rangle \langle \psi_{AB}^{(i)}| 
\end{align}
where the semiclassical baby universe may emerge as an effectively isolated region. 
A mixed state interpretation of a baby universe has also appeared in a recent work~\cite{Kudler-Flam:2025cki,Antonini:2025ioh}.
Our result provides an information theoretic proof of the mixed state interpretation. 

There, however, sill remain important conceptual puzzles. 
If a semiclassical baby universe possesses a large internal Hilbert-space dimension, it appears that its microstates are cloned. 
Namely, one copy lives as degrees of freedom within the baby universe itself, while the other is encoded in the entanglement patterns between two asymptotic AdS regions. 
This apparent cloning puzzle can be resolved by appealing to an \emph{emergent} form of complementarity, in which the baby-universe observer and the boundary observer have no operational means to compare or verify the cloning. 
The essential distinction we emphasize here is that, in our picture, complementarity is not assumed as a fundamental principle but rather arises dynamically from the structure of random encoding. 
Any communication between a single boundary observer and the baby-universe observer is prohibited as the baby-universe Hilbert space cannot be reconstructed from either boundary alone.

If we postulate that the baby universe and the asymptotic AdS regions are entangled through a tripartite Haar-like structure in the purified state $|\Psi_{ABC}\rangle$, we then face the question of how this picture can be reconciled with the apparent fate of the baby universe, which seems to eventually collapse into a singularity. 
This puzzle might be addressed by invoking a variant of the complementarity idea, possibly with additional observer degrees of freedom and non-isometric dynamics, or by hoping that full quantum gravity effects somehow allow information to escape from the baby-universe singularity. 
Although we do not attempt to resolve this issue here, we point out that the baby universe appears to be almost frozen from the boundary CFT viewpoint due to breakdown of complementary recovery. 
In particular, we argue that the effect of boundary time evolution generated by $H = H_A + H_B$ on the baby-universe degrees of freedom is exponentially suppressed scaling as $e^{-O(S_{\mathrm{BU}})}$. 

While our discussion is primarily motivated by two-sided geometries, it is natural to ask whether the same mechanism can account for baby-universe degrees of freedom in one-sided constructions. 
We extend our information-theoretic analysis to single-boundary setups by modeling the one-sided geometry as a projection of a tripartite Haar-random state. 
We demonstrate that the baby-universe remains operationally well-defined even after eliminating one boundary, in the sense that its entropy cannot be reduced by any local measurements on the discarded subsystem. 
This suggests that the emergence of baby-universe degrees of freedom does not fundamentally rely on the presence of two asymptotic boundaries.

This paper is organized as follows. 
Section~\ref{sec:tripartite}  reviews the typical absence of bipartite entanglement in tripartite Haar random states.
Section~\ref{sec:encoding} discusses the breakdown of complementary recovery in Haar random encoding and introduces a toy model of a baby universe. 
Section~\ref{sec:CFT} discusses how such pseudorandom tripartite entanglement structures may arise in genuine AdS/CFT setups. 
Section~\ref{sec:bulk} addresses conceptual issues in the bulk, including cloning and singularity puzzles. 
Section~\ref{sec:single} addresses the case of single-boundary baby universes. 
Finally, section~\ref{sec:outlook} summarizes our findings and reflecs on their conceptual implications. 

\section{Tripartite random state}\label{sec:tripartite}

In this section, we review the entanglement structure of tripartite Haar random state, focusing on the typical absence of bipartite entanglement~\cite{Li:2025nxv}.

We begin with an $n$-qubit Haar random state $|\Psi_{AB}\rangle$ supported on two complementary subsystems $A$ and $B$ with $n_A$ and $n_B = n-n_A$ qubits respectively. 
For $n_A < n_B$, we have 
\begin{align}
\mathbb{E}\  \Big\Vert \rho_A - \frac{1}{2^{n_A}}I_A \Big\Vert_1 \lesssim 2^{(n_A-n_B)/2},
\end{align}
where $\mathbb{E}$ represents the Haar average and $\frac{1}{2^{n_A}}I_A$ is the maximally mixed state on $A$~\cite{Page:1993df, Lubkin:1978nch, Lloyd:1988cn}.\footnote{Here, choosing a Haar random state means sampling uniformly from the set of all $n$-qubit pure states.
The Haar measure is the unique probability distribution that is left- and right-invariant under unitary transformations}
This implies that $A$ is nearly maximally entangled with a $2^{n_A}$-dimensional subspace in $B$. 
That is, there exists a local unitary $I_A \otimes U_B$ acting solely on $B$ such that
\begin{align}
(I_A \otimes U_B)|\Psi\rangle_{AB} \approx {|\text{EPR}\rangle^{\otimes n_A}}_{AA'}\otimes |\text{something}\rangle, \qquad |\text{EPR}\rangle = \frac{1}{\sqrt{2}}(|00\rangle + |11\rangle),
\end{align}
where $A'\subseteq B$ and $|A|=|A'|$, with $|R|$ representing the number of qubits in subsystem $R$. 
Hence, $n_A$ approximate EPR pairs can be distilled between $A$ and $B$ by applying some local unitary transformations without the need for measurements or classical communication. 
This result is central to understanding the heuristic behavior of evaporating black holes, forming the basis for the Page-curve intuition.

Next, let us extend this analysis to a tripartite $n$-qubit Haar random state $|\Psi_{ABC}\rangle$ defined on $A$, $B$, and $C$.
When one subsystem occupies more than half of the total system, the remaining subsystems are maximally entangled with it, effectively reducing the situation to the bipartite case. 
We therefore focus on the regime where each subsystem occupies less than half of the total, satisfying
\begin{align}
S_{R} \approx  n_R \qquad \big(0 < n_R < \frac{n}{2}\big) 
\end{align}
for $R=A,B,C$, where $S_R$ denotes the entanglement entropy. 
Two subsystems, say $A$ and $B$, exhibit a large amount of correlations, as seen in the mutual information:
\begin{align}
S_{AB} \approx  n_C, \qquad I(A:B) \equiv S_A + S_B - S_{AB} \approx n_{A} + n_{B} - n_{C} \sim O(n).
\end{align}
While mutual information does not by itself distinguish classical from quantum correlations, correlations in Haar random states are genuinely quantum. 
In particular, the logarithmic negativity~\cite{Lu:2020jza, Shapourian:2020mkc} satisfies
\begin{align}
E_{N}(A:B) \approx \frac{1}{2} I(A:B), \qquad E_{N}(A:B) \equiv \log_2 \Big( \sum_j |\lambda_j| \Big),
\end{align}
where $\lambda_j$ are the eigenvalues of the partial-transposed density matrix $\rho_{AB}^{T_{A}}$. 

A naturally arising question then concerns the nature of this quantum entanglement in $\rho_{AB}$.
In particular, can one distill EPR pairs from a single copy of $\rho_{AB}$ using some local unitary transformation $U_A\otimes U_B$, or more generally, local operation $\Phi_A\otimes \Phi_B$?~\footnote{
The entanglement distillation problem is conventionally discussed under LOCC (local operations with classical communication) and asks an asymptotic distillation rate from many copies of $\rho_{AB}$. 
Here, we focus on entanglement distillation from a single copy of $\rho_{AB}$ without classical communication. 
See~\cite{Mori:2024gwe} for 1WAY LOCC entanglement distillation from Haar random state and its holographic interpretation.
}

At first thought, one might expect that the large quantum correlations between $A$ and $B$ would allow the distillation of many EPR pairs.  
Indeed, for tripartite \emph{stabilizer} random states, the mixed state $\rho_{AB}$ contains $\approx \frac{1}{2}I(A:B)$ copies of unitarily (Clifford) rotated EPR pairs. 
This follows from the fact that tripartite stabilizer states admit only two types of entanglement: bipartite (EPR-like) and GHZ-like~\cite{PhysRevA.81.052302, Nezami:2016zni}, with the latter being rare in random stabilizer states~\cite{smith2006typical}.

Do tripartite Haar random states also consist mostly of bipartite entanglement? 
Recent results show that no EPR pairs can be distilled from $\rho_{AB}= \Tr_C \big(|\Psi_{ABC}\rangle \langle \Psi_{ABC}| \big)$ via local unitary transformations or local operations when $n_R < \frac{n}{2}$ for $R=A,B,C$. 

\begin{theorem}[informal]\label{theorem_ED}
The probability of sampling a quantum state $|\psi_{ABC}\rangle$ with locally distillable EPR pairs are exponentially suppressed with respect to $d$:
\begin{align}
\log \mathrm{Prob}(E_{D} \not= 0) \leq  - \alpha d  + O( d_A^2, d_B^2  )
\end{align}
where $\alpha>0$ is some constant and $E_{D}$ denotes the number of EPR pairs distillable by local unitaries or operations.
\end{theorem}

A precise version of theorem~\ref{theorem_ED}, including the definition of distillable entanglement with error tolerance, is presented in~\cite{Li:2025nxv} and is reprinted in appendix~\ref{sec:theorem}.
This result shows that tripartite Haar random states almost surely contain no bipartite entanglement.

One might then think that such states consist primarily of GHZ-like correlations.
However, an analogous argument rules out GHZ-type entanglement, and a large logarithmic negativity also rules out this possibility.\footnote{
This is also different from W-state and its generalizations as their local marginals are far from maximally mixed.
}
It is important to emphasize that ``$E_{D}=0$'' does not imply the absence of quantum correlation: tripartite Haar random states possess genuinely multipartite quantum entanglement that is neither bipartite nor classical. 

Strictly speaking, the absence of bipartite entanglement is an \emph{asymptotic} statement that becomes exact only in the large-$n$ limit. 
Nevertheless, the probability of successful distillation is suppressed \emph{doubly exponentially} with the number of qubits $n$, so the required scaling for the asymptotic behavior to emerge is remarkably mild.\footnote{Even though the baby-universe entropy $S_{\mathrm{BU}}$ may be subleading in $1/G_N$ below the Hawking-Page temperature, it typically scales polynomially in $1/G_N$ and can therefore be parametrically large~\cite{Antonini:2025ioh}.}

\section{Toy baby universe: Haar random encoding}\label{sec:encoding}

In this section, we demonstrate the breakdown of complementary recovery in Haar random encoding and interpret it as an analogue of the baby universe. 

Given a tripartite Haar random state $|\psi_{ABC}\rangle$, consider the bipartition into $AB$ and $C$. 
Since $C$ is (nearly) maximally entangled with a $2^{n_C}$-dimensional subspace of $AB$, we can express $|\psi_{ABC}\rangle$ with a Haar random isometry $V: C'\rightarrow AB$:
\begin{align}
|\psi_{ABC}\rangle \approx (V_{C'\rightarrow AB}\otimes I_{C}) |\text{EPR}\rangle_{C'C} = 
\figbox{2.0}{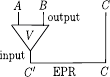} \ . 
\end{align}
In this diagram, the triangle denotes the isometry embedding an input state into a larger Hilbert space, and the horizontal line between $C$ and $C'$ represents EPR pairs. 
This construction allows us to view a Haar random state $|\Psi_{ABC}\rangle$ as defining a random quantum error-correcting code $V$ that encodes $k=n_C$ logical qubits into $n_A + n_B$ qubits.
This dual interpretation is often referred to as the channel-state duality, and it plays a central role in explaining the physical mechanisms underlying black hole information recovery~\cite{Hosur:2015ylk}. 

We now ask whether the encoded quantum information is recoverable from a single subsystem, say $A$, when the complementary subsystem $B$ is lost (e.g., due to erasure).
This is equivalent to asking whether a \emph{logical operator} can be supported entirely on $A$.
Loosely speaking, $\overline{U}$ is a logical representation of $U$ if it reproduces the logical action of $U$ within the code subspace:
\begin{align}
\figbox{2.0}{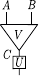} \ = \ \figbox{2.0}{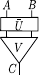}
\end{align}
If a logical operator can be supported entirely on $A$, then the subsystem $A$ retains non-trivial quantum information about $C$. 

The relation to the entanglement distillation problem becomes evident by considering a pair of anti-commuting Pauli logical operators. 
If Pauli logical operators $\overline{X}, \overline{Z}$ could both be supported on $A$, then an EPR pair could be distilled between $A$ and $C$ in $|\Psi_{ABC}\rangle$. 
Since theorem~\ref{theorem_ED} established that no EPR pairs can be distilled in tripartite Haar random state, we conclude that no such logical Pauli operators can be supported inside $A$. 
This observation can be extended and formalized as the following no-go theorem. 

\begin{theorem}[informal]\label{theorem_logical}
    Consider a Haar random encoding $C\to AB$. If $n_A<n_B+n_C$ and $n_B<n_A+n_C$, then neither $A$ nor $B$ contain any quantum information about $C$.
    Equivalently, no nontrivial logical unitary operator can be represented as an operator supported entirely on $A$ or on $B$.  
\end{theorem}

A precise statement of theorem~\ref{theorem_logical} is presented in~\cite{Li:2025nxv} and is reprinted in appendix~\ref{sec:theorem}.
This theorem demonstrates a complete breakdown of complementary recovery in Haar random encoding. 

At first sight, this may be counterintuitive: even though the input $C$ and the output $A$ have large quantum correlation with $I(A:C)\sim O(n)$, this does not grant power to reconstruct logical quantum information of the code. 
One crucial point is that this is an asymptotic statement where the ``quality'' of logical operator reconstruction becomes exponentially bad with respect $d$. 
Namely, the correlation between $A$ and $C$ is not sufficient to guarantee high fidelity logical operator reconstruction.
\footnote{ 
We note that this statement goes beyond what is currently understood from entanglement wedge considerations. In certain semiclassical gravitational setups, the union of entanglement wedges associated with two boundaries does not appear to cover the entire bulk~\cite{Goel:2018ubv,Iizuka:2021tut,Balasubramanian:2023xyd,Antonini:2025ioh} (see also~\cite{Marolf:2020xie}). However, the precise boundary operator interpretation of such situations remains subtle~\cite{Akers:2020pmf,Akers:2023fqr}.
}

One might think that logical operators can then be written as a non-local, but factorized form over $A$ and $B$, $\overline{U}= \overline{U}_A\otimes \overline{U}_B$. 
However, an analogous argument also rules out the possibilities of factorized non-local logical operators, suggesting $\overline{U}\not= \overline{U}_A\otimes \overline{U}_B$.


Here it is essential that the logical operators $\overline{U}$ are \emph{unitary}. 
Indeed, certain \emph{non-unitary} logical operators may still be reconstructed on $A$ or $B$, depending on the sizes of $A$ and $B$. 
To illustrate this, let us further decompose the input system as $C=C_{0}\otimes C_1$ and consider the operator $V = X_{C_0} \otimes |0\rangle\langle 0 |_{C_{1}}$ 
which includes a projection operator on subsystem $C_1$. 
This projection effectively reduces the logical input to the smaller subsystem $C_0$. 
If $A$ then occupies more than half of the total system $ABC_0$, the logical operator $\overline{X_{C_0}}$ can be supported entirely on $A$. 
Because the explicit form of such logical operators depends on the chosen projected state $|0\rangle\langle 0|_{C_1}$, this constitutes an example of state-specific reconstruction~\cite{Hayden:2018khn, Yoshida:2017non}. 
In the entanglement distillation problem, this corresponds to one-way LOCC distillation~\cite{Mori:2024gwe}, where the projection $|0\rangle\langle0|_{C_{0}}$ is implemented by local projective measurements $\{ |i\rangle\langle i| \}$ followed by classical communication. 

To appreciate how striking this property is, let us contrast it with a familiar example, the GHZ state (or equivalently, the classical repetition code) which \emph{fails} to exhibit a full breakdown of complementary recovery.
The classical repetition code encodes a single bit of information into repeated bits: $0 \mapsto 00$ and $1 \mapsto 11$. 
Extending this to the quantum setting defines the isometric map 
\begin{align}
\alpha|0\rangle + \beta|1\rangle \rightarrow \alpha|00\rangle + \beta|11\rangle
\end{align}
which encodes one qubit from $C$ into two physical qubits in $A$ and $B$. 
Expressing this isometry as a tripartite pure state ($|00\rangle_{AB}\otimes |0\rangle_C + |11\rangle_{AB}\otimes |1\rangle_C$)  yields the GHZ state:
\begin{align}
|\text{GHZ}\rangle = \frac{1}{\sqrt{2}}( |000\rangle + |111\rangle).
\end{align}

We can now ask whether logical operators can be supported locally on $A$ or $B$. 
One can observe that 
\begin{align}
&(I \otimes I \otimes Z_C) |\text{GHZ}\rangle = (Z_A \otimes I \otimes I) |\text{GHZ}\rangle = (I \otimes Z_B \otimes I) |\text{GHZ}\rangle \\
&(I \otimes I \otimes X_C) |\text{GHZ}\rangle = (X_A \otimes X_B \otimes I) |\text{GHZ}\rangle.
\end{align}
Thus, the logical operators can be written as 
\begin{align}
\overline{X} = X_A \otimes X_B,\qquad \overline{Z} = Z_A \otimes I_B \sim I_A \otimes Z_B.
\end{align}
Hence the GHZ state achieves only a partial breakdown of complementary recovery. Namely, the logical Pauli $X$ cannot be realized locally on $A$ or $B$, whereas the logical Pauli $Z$ can be reconstructed from either side. 

It is instructive to contrast Haar random encoding with random Clifford encoding. 
When $|\Psi_{ABC}\rangle$ is a tripartite random stabilizer state, or equivalently when $V$ is a random Clifford isometry, the encoded logical qubits can be locally recovered from subsystems $A$ and $B$.
Specifically, there exists a Clifford operator $U_0$ acting on $C$ such that 
\begin{align}
\figbox{2.0}{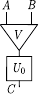} \ = \ \figbox{2.0}{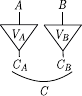} \qquad \text{(random Clifford encoding)}
\end{align}
In other words, one can reorganize the input system as $C\simeq C_A \otimes C_B $, so that $C_A$ ($C_B$) can be reconstructed from $A$ ($B$). 
This follows from the fact that tripartite stabilizer random states consist predominantly of bipartite entanglement. 
Consequently, complementary recovery holds exactly for random stabilizer encoding with overwhelming probability.

We now take a step further and interpret a Haar random isometry as an analogue of a baby universe. 
In AdS/CFT, bulk quantum gravity degrees of freedom are holographically encoded into boundary quantum systems in a way fundamentally akin to a quantum error-correcting code.
The conceptual pillar behind this picture is \emph{entanglement wedge reconstruction}, which asserts that if a bulk operator $\phi$ lies within the entanglement wedge $\mathcal{E}_A$ of a boundary subsystem $A$, then it can be represented as a boundary operator $O_A$ supported entirely on $A$.
 
\begin{figure}
\centering
\includegraphics[width=0.35\textwidth]{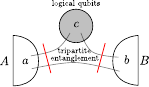}
\caption{A baby universe appears as an effectively closed region outside entanglement wedges of two individual boundaries.
}
\label{fig_baby_cartoon}
\end{figure}

In the present analogy, we interpret the input subsystem $C$ as the bulk degrees of freedom, while the output subsystems $A$ and $B$ represent two boundary CFTs:
\begin{align}
\figbox{2.0}{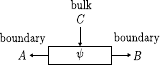}  
\end{align}
The minimal surfaces associated with $A$ and $B$ are schematically depicted as
\begin{align}
S_A = 
\figbox{2.0}{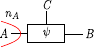}\ , \qquad  S_B = 
\figbox{2.0}{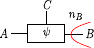}
\end{align}
where the bulk region $C$ lies outside both entanglement wedges $\mathcal{E}_A$ and $\mathcal{E}_B$. 
Can either boundary reconstruct bulk operators on $C$? 
Theorem~\ref{theorem_logical} states that no bulk unitary operator can be supported on $A$ or $B$. 
Therefore, in Haar random encoding, bulk operators lying outside the entanglement wedge are not reconstructable, precisely mirroring the expected behavior of semiclassical baby universes. 
From the viewpoint of either boundary, the Hilbert space $C$ appears absent; yet it exists as logical qubits accessible only through non-local joint operations acting on $A$ and $B$.
In this sense, a random encoding realizes a baby-universe-like sector emerging as an effectively closed region invisible to each individual boundary observer (Fig.~\ref{fig_baby_cartoon}). 

In the language of quantum error correction, the baby universe corresponds to the logical qubits of the code, with each codeword representing a distinct baby-universe microstate.
In principle, the two boundary CFTs could jointly reconstruct these logical states, but only through nonlocal operations.
From the perspective of two causally disconnected boundaries, observers on $A$ and $B$ cannot access the detailed structure of the codeword states individually. 
Operationally, the only thing $A$ and $B$ can confirm is that their joint state is supported on the code subspace of a particular isometry $V$. 
Under this condition, they would describe their joint system by a mixed state
\begin{align}
\rho_{AB} = \Tr_C \big( |\Psi_{ABC}\rangle\langle \Psi_{ABC}| \big)
\end{align}
which is precisely the maximal entropy state within the code subspace. 
This $\rho_{AB}$ represents their best estimate using local boundary algebras, assuming knowledge of the encoding map $V$, but not of the input state on $C$.
\footnote{ 
If they do not even know $V$, their description further reduces to a product of maximally mixed states, $\rho_A \otimes \rho_B \approx \frac{1}{d_A d_B} I_A \otimes I_B$. 
}
From this boundary perspective, the baby-universe entropy $S_{\mathrm{BU}}$ can be identified with the entropy of the emergent mixed state $\rho_{AB}$. 
Equivalently, $S_{\mathrm{BU}}$ may be interpreted as the coarse-grained entropy associated with restricting to the local boundary algebra, conditioned on knowing the associated code subspace. 


At this point, it is useful to recall how the Ryu-Takayanagi (RT) formula relates to complementary recovery. 
As shown in~\cite{Harlow:2016vwg}, within the von Neumann algebra framework, the RT relations for complementary boundary subsystems $A$ and $B$ hold \emph{if and only if} complementary recovery ($a \mapsto A$ and $b\mapsto B$) is valid. 
At first glance, theorem~\ref{theorem_logical} then might seem to imply a breakdown of the RT formula. 
However, there is no contradiction: the argument of~\cite{Harlow:2016vwg} presumes RT relations of the form
\begin{align}
S_{A} \approx \Tr( \rho_a L_a  ) + S_{\text{bulk}}(\rho_a ), \qquad 
S_{B} \approx \Tr( \rho_b L_b  ) + S_{\text{bulk}}(\rho_b )
\end{align}
with ``area operators'' $L_a$ and $L_{b}$ are identified as a shared central element of the algebras associated with bulk regions $a$ and $b$ ($L_a= L_b$). 
Theorem~\ref{theorem_logical} simply points to the fact that this identification may not hold, $L_a \not= L_{b}$, leaving a residual bulk region lying outside both entanglement wedges $\mathcal{E}_A$ and $\mathcal{E}_B$. 


\section{Baby universe in CFTs}\label{sec:CFT}

Let us begin by recalling the semiclassical baby universe geometry constructed by AS$^2$~\cite{Antonini:2023hdh}.
We consider a setup where two CFTs $A$ and $B$ are prepared in a thermal state below the Hawking-Page transition, with the insertion of a shell of heavy Euclidean operators $O$:\footnote{In principle, the inverse temperatures on the left $\beta_L$ and the right $\beta_R$ of the operators can be different. Since it does not affect the following discussion, we focus on the case when $\beta_L=\beta_R=\beta$.}
\begin{align}
|\Psi_{AB}\rangle = \frac{1}{\sqrt{Z}} \sum_{m,m'} e^{- \frac{1}{2}\beta(E_m + E_{m'})}O_{mm'} |E_m\rangle |E_{m'}\rangle.
\end{align}
To avoid contributions from black-hole saddles, we may further restrict to energy levels strictly below the Hawking-Page temperature.\footnote{Each heavy operator has a conformal dimension $1\ll \Delta \ll \Delta_{\rm Hawking-Page}=O(c)$.} 

\begin{figure}
\centering
\includegraphics[width=0.45\textwidth]{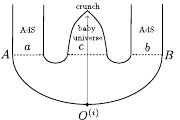}
\caption{Semiclassical baby universe. 
}
\label{fig_semiclassical_baby}
\end{figure}

Naively, such a boundary state should correspond to two disconnected asymptotically AdS spacetimes
$a$ and $b$ where the two CFTs are entangled only through bulk matter fields. 
However, AS$^2$ showed that the dominant gravitational saddle of this path integral includes a third closed region $c$ in the bulk (Fig.~\ref{fig_semiclassical_baby}). 
This baby universe exists for a finite time before contracting into a singularity (``crunch''). 
Moreover, the heavy operator $O$ inserted in the Euclidean path integral can propagate into the closed region, suggesting that the baby universe supports semiclassical trajectories and local dynamics. 

The appearance of a semiclassical baby universe is highly non-trivial.
While, baby universes have long been invoked in discussions of the AdS/CFT factorization problem~\cite{Marolf:2020xie}, those construction involved non-semiclassical or topological sectors. 
In contrast, the geometry here is smooth and seemingly accessible to measurement probes.  
However, a closed universe without asymptotic boundaries should have a one-dimensional Hilbert space at the semiclassical level. 
 
Antonini and Rath (AR) further sharpened the puzzle~\cite{Antonini:2024mci}.
The path-integral geometry consists of three disconnected regions $a,b,c$, and a bulk state $|\psi^{(1)}_{abc}\rangle$ where $a,b$ correspond to two asymptotic AdS regions and $c$ represents the baby universe. 
They then rewrote the same boundary state using local, low-dimensional boundary operators $\Phi_A^i, \Phi_B^j$ acting on the vacuum,
\begin{align}
|\Psi_{AB}\rangle \approx \sum_{A_i, B_j} c_{A_i, B_j} |A_{i}\rangle |B_j\rangle
\end{align}
where $ |A_{i}\rangle=\Phi_A^i|0\rangle_A$ and $|B_j\rangle=\Phi_B^j|0\rangle_B$. 
Using the standard holographic encoding maps (i.e. the HKLL reconstruction~\cite{Hamilton:2005ju}):
\begin{align}
W_a : \mathcal{H}_a \rightarrow \mathcal{H}_A, \quad W_b : \mathcal{H}_b \rightarrow \mathcal{H}_B,
\end{align}
the full encoding is $W = W_{a}\otimes W_{b}$, leading to
\begin{align}
|\Psi_{AB}\rangle = W\Big(  \sum_{i,j}\alpha_{A_i, B_j} |a_{i}\rangle |b_j\rangle  \Big).
\end{align}
So, $|\Psi_{AB}\rangle$ is dual to a single pure bulk state $|\psi^{(2)}_{ab}\rangle$ on $ab$, namely 
\begin{align}
|\psi^{(2)}_{ab}\rangle =  \sum_{i,j}\alpha_{A_i, B_j} |a_{i}\rangle |b_j\rangle.
\end{align}
Hence, the same boundary state $|\Psi_{AB}\rangle$ appears to admit two distinct semiclassical bulk interpretations. 

Here we consider an ensemble of randomly chosen heavy operators $\mathcal{M}_{\mathrm{heavy}} = \{O^{(i)}\}_{i=1}^M$, and study the corresponding ensemble of boundary states. 
While $M$ may be taken arbitrarily large in principle, the effective number of independent heavy operators is bounded due to the sparseness of low-energy excitations in holographic CFTs. 
We will come back to this point later in this section. 

Each insertion constructs a two-sided CFT wavefunction expanded in the low-energy basis below the Hawking-Page temperature as
\begin{align}
|\Psi_{AB}^{(i)} \rangle = \frac{1}{\sqrt{Z^{(i)}}} \sum_{m,n} \tilde{O}^{(i)}_{mn}|m\rangle_{A} \otimes |n\rangle_{B} = \figbox{2.0}{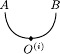} 
\label{eq:tilde-O-pets}
\end{align}
where $\tilde{O}^{(i)}_{mn}$ are matrix elements of the Euclidean-evolved heavy operator. 
By construction, the state $|\Psi_{AB}^{(i)}\rangle$ is predominantly supported in the low-energy subspaces of the two CFTs, below the Hawking-Page transition temperature. 

Our working hypothesis is that the tensor $\tilde{O}^{(i)}_{m,n}$ exhibits pseudorandom structure in a sense that its fine-grained statistics resemble those of tripartite Haar random states. 
This assumption is admittedly speculative, but it is motivated by several hints from holographic CFTs. 
Correlation functions that mix heavy and light operators are expected to display significant randomness, similar to the eigenstate thermalization hypothesis (ETH) behavior~\cite{Lashkari:2016vgj, Collier:2019weq, Belin:2020hea,Anous:2021caj,Belin:2023efa}, once we move beyond their smooth semiclassical averages in the large $N$ limit. 
Moreover, the Euclidean evolution $e^{-\frac{\beta H}{2}}$ acts with a strongly chaotic CFT Hamiltonian, which further washes out detailed structure and enhances pseudorandom features. 

To make the tripartite structure more explicit, it is convenient to introduce a fictitious third boundary system $C$ that records the label $i$ of the heavy  operator:
\begin{align}
|\Psi_{ABC}\rangle = \frac{1}{\sqrt{M}} \sum_{i=1}^{M}|\Psi_{AB}^{(i)}\rangle \otimes |i\rangle_{C}= 
\frac{1}{\sqrt{Z}} \sum_{i,m,n} \tilde{O}^{(i)}_{mn}|m\rangle_{A} \otimes |n\rangle_{B} \otimes |i\rangle_C.
= \figbox{2.0}{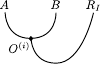} \ .
\end{align}
If $\tilde{O}^{(i)}_{mn}$ indeed behaves pseudorandomly, then the state $|\Psi_{ABC}\rangle$ can be viewed as an approximate tripartite random state over $A$, $B$, and $C$. 
In this picture, the detailed information about the initial heavy operator $O^{(i)}$ is encoded nonlocally in the joint system $AB$. 

One remaining question concerns how to construct the ensemble of heavy operators $\{ O^{(i)} \}_{i=1}^M$.
For our purposes, we simply sample heavy operators with similar macroscopic quantum numbers uniformly at random. 
Because of their inherent pseudorandomenss, two boundary states $|\Psi_{AB}^{(i)}\rangle$ and $|\Psi_{AB}^{(j)}\rangle$ are expected to be nearly orthogonal when the corresponding heavy operators are chosen independently.
If we treat the states $|\Psi_{AB}^{(i)}\rangle$ as approximately random within some subspace, this suggests that the entropy of the reference system $C$ in the purified state $|\Psi_{ABC}\rangle$ should grow as $S_C \sim \log M$.
However, this growth cannot continue indefinitely as $|\Psi_{ABC}\rangle$ is a pure state, so $S_C \leq S_A + S_B$.
Thus, the entropy $S_C$ must eventually saturate. 
In regimes where a semiclassical baby universe exists, boundary entropies $S_A, S_B$ scales polynomially with $N$~\cite{Antonini:2025ioh}. 

We will assume that this saturation occurs at a scale set by the effective baby-universe entropy $S_{\mathrm{BU}}$.
Heuristically, we expect the relation
\begin{equation}
\begin{split}
S_C &\approx \log M \qquad M \lesssim e^{S_{\text{BU}}} \\ 
&\approx S_{\text{BU}} \qquad  \ \ M \gtrsim e^{S_{\text{BU}}}.
\end{split}
\label{eq:BU-Page}
\end{equation} 
This relation is partly guided by the bulk geometry, in which the semiclassical baby universe appears to carry an extensive entropy and can trap information about the initial heavy operator insertions.
However, we stress that \eqref{eq:BU-Page} is formulated purely as a boundary characterization of the baby-universe entropy $S_{\mathrm{BU}}$, without invoking any detailed bulk construction. 
\footnote{
Here, the effective baby-universe entropy $S_{\mathrm{BU}}$ has been defined purely from a boundary perspective. 
A subtle issue is that the baby-universe entropy $S_{\mathrm{BU}}$ can be different from the bulk thermodynamic entropy $S_{\mathrm{BU}}^{(\mathrm{th})}$ of the closed region computed. 
Depending on $\beta$ and $\Delta$, it is possible that $S_{\mathrm{BU}}^{(\mathrm{th})}$ is parametrically larger than $S_A, S_B$. 
In such regimes, we necessarily have $S_{\mathrm{BU}}^{(\mathrm{th})}\gg S_{\mathrm{BU}}$ since $S_{\mathrm{BU}}$ is bounded by $S_A + S_B$. 
This raises an important interpretational question: does the boundary CFT capture only an $e^{S_{\mathrm{BU}}}$-dimensional subspace of a potentially larger baby-universe Hilbert space, or should we regard only that $e^{S_{\mathrm{BU}}}$-dimensional subspace as physically admissible in light of holographic constraints? 
At present, both interpretations are logically consistent, and we do not take a position here. 
A recent analysis~\cite{Antonini:2025ioh} estimates $S_{\mathrm{BU}}^{(\mathrm{th})}$ and emphasizes the operational distinction between $S_{\mathrm{BU}}$ and $S_{\mathrm{BU}}^{(\mathrm{th})}$.
}


From the viewpoint of \eqref{eq:BU-Page}, it is not necessary to include all possible heavy operators. 
Instead, one may restrict to a subset of size $M\sim e^{S_{\mathrm{BU}}}$ chosen at random. 
A related viewpoint is that the Euclidean operators $\tilde{O}^{(i)}$ involve non-unitary evolution, and thus only a coarse-grained imprint of the original heavy operators $O^{(i)}$ survives in the state $|\Psi_{ABC}\rangle$. 
Consequently, the effective dimension of distinguishable heavy operators is of order $e^{S_{\mathrm{BU}}}$, which may be interpreted as the boundary manifestation of the baby-universe entropy. 
Careful readers will notice that this coarse-graining reflects a form of non-isometricity~\cite{Akers:2022qdl}. We will return to this point later.

Finally, by tracing out the reference system $C$, we obtain the mixed state
\begin{align}
\rho_{AB} = \frac{1}{\sqrt{M}} \sum_{i=1}^{M} |\Psi_{AB}^{(i)}\rangle \langle \Psi_{AB}^{(i)}| = \figbox{2.0}{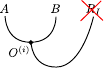} 
\end{align} 
which reflects the fact that boundary observers on $A$ and $B$ cannot access the microscopic structure of the initial heavy operators individually. 

The emergence of this mixed state by ``forgetting'' the label $i$ reminds us of a familiar feature of gravitational path integrals, coarse-graining over microscopic input data often leads to averaged or ensemble-like quantities~\cite{Marolf:2020xie,deBoer:2023vsm}.
However, the present picture is stronger than coarse-graining over a classical label. 
The indistinguishability we obtain here is not merely about recovering which $O^{(i)}$ was inserted, as it extends to superpositions of heavy operators. 
In other words, both classical and quantum information about the initial insertion are inaccessible to any single boundary. 

It is useful to highlight the classical v.s. quantum distinction. 
Classical indistinguishability can be achieved in many ways, for example via GHZ-like entanglement. 
But such constructions still allow quantum information about superpositions to be recovered from each boundary individually. 
Consider a simple 
three-qubit system on $A$, $B$, and $C$, with
\begin{align}
\tilde{O}^{(0)} = I = |0\rangle\langle0| +|1\rangle\langle1|, \qquad \tilde{O}^{(1)} =X = |0\rangle\langle1| +|1\rangle\langle0|.
\end{align}
This leads to the tripartite state~\eqref{eq:tilde-O-pets}
\begin{align}
|\Psi_{ABC}\rangle = \frac{1}{2} \big(|0\rangle |0\rangle |0\rangle +|1\rangle |1\rangle |0\rangle +|0\rangle |1\rangle |1\rangle +|1\rangle |0\rangle |1\rangle  \big)
\end{align}
which satisfies
\begin{align}
&(Z_A\otimes Z_B \otimes Z_C)|\Psi_{ABC}\rangle = |\Psi_{ABC}\rangle \\
&(X_A\otimes I \otimes X_C)|\Psi_{ABC}\rangle = (I\otimes X_B \otimes X_C)|\Psi_{ABC}\rangle= |\Psi_{ABC}\rangle
\end{align}
Thus, the label $i$, encoded in the eigenvalues of $Z_C$, is not recoverable from $A$ or $B$ individually.
However, if we project $C$ onto $|0\rangle \pm |1\rangle$ (corresponding to inserting $O^{(0)}\pm O^{(1)}$), then the relative $\pm$ phase, encoded in the eigenvalues of $X_C$, is reconstructable locally on both $A$ and $B$. 
This illustrates why GHZ-type entanglement is insufficient for capturing essential baby-universe property. 
 
\section{Bulk picture: cloning puzzle and singularity}\label{sec:bulk}

Our information-theoretic analysis has been based on the assumption that the semiclassical bulk may contain the baby-universe-like object with an effective entropy $S_{\mathrm{BU}}$. 
We are then led to view the baby universe as emergent degrees of freedom, analogous to logical qubits in a random quantum error-correcting code, which escape complementary recovery due to the inherent pseudorandomness in holographic CFTs. 
We now turn to the question of whether such a picture is consistent from the bulk point of view. 

Two important issues must be addressed here: a potential cloning of baby-universe microstates, and the singularity of the closed region. 
We first analyze the $t=0$ slice and postpone discussion of the singularity. 
Let the two asymptotic AdS regions be denoted by $a$ and $b$, and let the closed region be $c$. 
For a fixed heavy operator $O^{(i)}$, the boundary state $|\Psi_{AB}^{(i)}\rangle$ lies entirely in the low-energy subspace of two CFTs, and therefore corresponds semiclassically to two AdS regions $a,b$ entangled via bulk matter fields. 
If a baby universe $c$ exists at $t=0$ in addition to two AdS regions, its state must be fully decoupled from the asymptotic regions $a,b$:
\begin{align}
|\psi_{abc}^{(i)}\rangle = |\psi_{ab}^{(i)}\rangle \otimes |\psi_{c}^{(i)}\rangle
\end{align}
where $|\psi_{c}^{(i)}\rangle$ denotes the bulk state of the closed region $c$ that depends on the choice of heavy operator. 
This factorization is consistent with the fact that the baby universe has no boundary, and therefore the RT formula would assign $S_c = 0$. 


\begin{figure}
\centering
\includegraphics[width=0.45\textwidth]{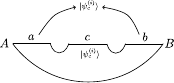}
\caption{A cloning puzzle at the $t=0$ slice. A copy of the bulk baby-universe state $|\psi_c^{(i)}\rangle$ can be reconstructed from two AdS regions $a,b$.
}
\label{fig_baby_cloning}
\end{figure}

However, this raises an apparent cloning puzzle: the bulk baby-universe state $|\psi_c^{(i)}\rangle$ is already encoded in $|\psi_{ab}^{(i)}\rangle$ as the entanglement structure of the asymptotic regions, so a bulk (or boundary) operation acting only on $a,b$ can in principle reconstruct $|\psi_c^{(i)}\rangle$. 
Thus, two copies of the same quantum state $|\psi_c^{(i)}\rangle$ appear in the $t=0$ slice. 
Equivalently, in the fully purified state $|\Psi_{ABC}\rangle$, the corresponding bulk state takes form  
\begin{align}
|\psi_{abc}\rangle = \frac{1}{\sqrt{M}}\sum_i |\psi_{ab}^{(i)}\rangle \otimes |\psi_{c}^{(i)}\rangle.
\end{align}
Since $c$ carries entropy $S_c\sim S_{\mathrm{BU}}$, it is nearly maximally entangled with $ab$, yet the label system $C$ is also entangled with $c$. 
This would seem to contradict the monogamy of entanglement. 

A naive way to avoid this puzzle is to take the baby-universe Hilbert space to be one-dimensional, so that there would be no non-trivial quantum information to be cloned, as advocated in e.g.~\cite{Engelhardt:2025vsp}. 
However, within a random-encoding picture, the apparent cloning tension can be alleviated by an emergent version of complementarity.
In particular, boundary observers on $A,B$ cannot reconstruct operators on $c$ locally due to the breakdown of complementary recovery.
The random encoding ensures that the baby-universe Hilbert space is inaccessible from either boundary alone, and therefore no physical protocol exists that would reveal a cloning of quantum information.
In this sense, the baby universe may carry a nontrivial entropy $S_{\mathrm{BU}}$ without contradicting the no-cloning or monogamy constraints. 
The key point is that complementarity emerges dynamically from the structure of the encoding rather than being imposed as an axiom.

This emergent complementarity suggests that a semiclassical baby universe can be reconciled with the boundary description as long as boundary operations are restricted to local ones.
A natural next question is what happens if the two boundary observers $A$ and $B$ are allowed to perform joint, non-local operations. 
We do not attempt to give a definitive answer here, but it is instructive to recall an analogous situation in the two-sided AdS black hole. 
There, quantum information theoretic arguments show that information falling behind the horizon can, in principle, be reconstructed if one couples the two boundary CFTs in an appropriate way, thanks to the scrambling dynamics~\cite{Hosur:2015ylk}. 
This immediately raises the question of how such a recovery is compatible with the geometric fact that the information lies behind the horizon. 
Several proposals exist, but one resolution particularly aligned with our viewpoint is that the recovery operation itself is highly complex and induces substantial backreaction on the semiclassical spacetime. 
Under this interpretation, the backreaction dynamically shifts the location of the horizon, effectively pulling the infalling information to the exterior. 
Although this may seem dramatic, a concrete realization of this mechanism already exists in the traversable-wormhole protocol, where a suitable boundary couplings makes behind-the-horizon information accessible~\cite{Gao:2016bin, Maldacena:2017axo}. 

In analogy with this, one may speculate that sufficiently complex joint operations on $AB$ could induce large backreaction on the semiclassical baby-universe geometry as well. 
Such operations might ``open'' the closed region and make it accessible to the boundaries. 
From the boundary perspective, this would correspond to implementing an appropriate recovery channel, such as a unitary realization of the Petz map or the double trace deformation.
Whether this interpretation is literally realized in the bulk remains open, but it provides a plausible framework for understanding how joint boundary actions might evade the cloning puzzle without contradicting semiclassical expectations. 

Next, let us turn to another important issue concerning the fate of the singularity in the baby universe. 
In the fully purified state $|\Psi_{ABC}\rangle$, the label system $C$ is nearly maximally entangled with the boundary degrees of freedom $AB$.
As long as the boundary evolves unitarily under $H=H_A + H_B$, the entanglement between $C$ and $AB$ remain intact.
On the other hand, the semiclassical bulk picture suggests that the heavy-operator excitations encoded in $C$ fall into the closed region $c$ and eventually encounter a singularity where they appear to be destroyed. 
This raises a familiar tension on how the boundary evolution preserves quantum information while the semiclassical bulk seem to destroy it. 

\begin{figure}
\centering
\includegraphics[width=0.55\textwidth]{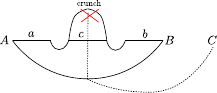}
\caption{A singularity puzzle. The reference system $C$ is initially entangled with heavy operators $O^{(i)}$ which  eventually fall into a singularity. Entanglement between $c$ and $C$ appears to be lost. 
}
\label{fig_baby_singularity}
\end{figure}


A crucial point is that, as long as the boundary evolves strictly under $H = H_A + H_B$, the baby-universe Hilbert space remains operationally inaccessible to either boundary subsystem. 
Because no boundary observer can reconstruct any nontrivial unitary operator on $C$, boundary time evolution does not translate into ordinary Hamiltonian evolution within the baby-universe sector. 
This effectively means that the degrees of freedom associated with the baby universe are frozen from the viewpoint of the boundary. 
This suggests that, at least at the level of effective dynamics probed by boundary local operations, the CFT may not directly witness the semiclassical collapse of the closed region. 

One can make this reasons slightly more quantitative. 
Recall that theorem~\ref{theorem_logical} is an asymptotic statement, asserting that reconstruction of logical operators on $C$ becomes impossible in the large $n$ limit.
Examining the probability bound shows that the fidelity of reconstruction decays as $\sim O(1/\sqrt{d_C})$. 
Hence, noticeable dynamics in the baby universe would require the boundary timescale of order 
\begin{align}
t \sim \exp(O(S_{\mathrm{BU}})).
\end{align}
In other words, from the CFT viewpoint, the baby-universe sector may evolve only on exponentially long timescales. 
Beyond such times, it is unclear whether semiclassical bulk intuition, or even the AdS/CFT correspondence itself, should be expected to apply.~\footnote{For example, the spectral form factor is expected to behave differently at exponentially late time in quantum gravity whose spectrum is discrete~\cite{Cotler:2016fpe}.}

\section{Single-boundary baby universe}\label{sec:single}

While most existing constructions of semiclassical baby universes arise in two-sided geometries, there is no fundamental reason why the existence of the baby-universe degrees of freedom should depend on the presence of two asymptotic boundaries. 
Moreover, one-sided geometries are closer to the settings encountered in dynamical gravitational processes, including the endpoint of black hole evaporation. 
Since our proposal relies crucially on the presence of two boundary systems ($A,B$), it is natural to ask how the picture changes in a one-sided construction. 

In this section, we present a brief discussion on potential implications to single-boundary baby universes.
A corresponding boundary state describing a single-boundary baby universe can be prepared by evolving the boundary from $t=-\infty$ to $t=0$, inserting a heavy operator, and then evolving in Euclidean time up to $\tau=\beta/2$~\cite{Kudler-Flam:2025cki, Balasubramanian:2025zey}.
Semiclassically, this procedure produces a baby universe $(C)$ entangled with a single asymptotic AdS region $(A)$ (but without the other AdS region $B$).
To model this situation in our toy framework, we begin with the tripartite Haar-random setup $(A,B,C)$ and obtain an effective bipartite system $(A,C)$ by projecting a subsystem $B$ onto some state.
We interpret this toy one-sided construction as projecting one boundary ($B$) onto a typical energy eigenstate.
We then ask whether this projection destroys the baby-universe degrees of freedom in $C$, or whether they remain as effectively independent degrees of freedom.

An information theoretic way to probe this question is through the \emph{locally accessible information} $J(C|B)$, which quantifies how much the entropy of $C$ can be reduced by measurements performed only on $B$. 
More precisely, for all possible measurements described by positive operator-valued measures (POVMs) $\{ \Pi^{i}_B \}_i$ on $B$, and the resulting reduced states $\{p_j, \rho_C^{j}  \}_j$, we define
\begin{align}
J(C|B) &\equiv S(\rho_C) - \min_{\{ \Pi^{i}_B \}_i} \sum_{j}p_j S_{C}(\rho_C^j), \qquad 
p_j = \Tr (\Pi^j_B \rho_B)
\end{align}
where $S(\rho_C)$ is the von Neumann entropy of the reduced state on $C$ before the measurement.
In the holographic context, $J(C|B)$ quantifies the number of EPR pairs that can be distilled between $C$ and $B$ using one-way LOCC operations~\cite{Mori:2024gwe} where classical communication is sent from $B$ to $C$ (but not the other way). 

Conceptually, the key question is whether the baby universe can carry independent degrees of freedom with extensive entropy and internal physics. 
If the entropy drop $J(C|B)$ is nearly zero, then no measurement of the discarded boundary significantly reduce the entropy of $C$, and the baby-universe degrees of freedom remain operationally well-defined as an effectively $e^{S_{C}}$-dimensional Hilbert space from the perspective of the remaining boundary system. 

For Haar random states, $J(C|B)$ has been computed previously~\cite{Hayden_2006} and is approximately given by
\begin{align}
J(C|B) \approx n_C - \min (n_A , n_C).
\end{align}
In particular, when $n_A > n_C$, this quantity is nearly zero.
This is a strong statement: regardless of how cleverly one chooses the measurement basis on $B$, the entropy of $C$ cannot be substantially reduced.
Importantly, this result makes no assumptions about the complexity of the measurement.
It suggests that the baby universe $C$ in the one-sided geometry remains well-defined as long as $S_A > S_C$ in the AS$^2$-type construction~\cite{Antonini:2023hdh}.  
Although there is no a priori reason to expect this inequality to hold universally, there exists a regime in which the information content of $C$ is not significantly affected by passing from the two-sided to the one-sided construction.

This, however, raises a tension with entanglement wedge reconstruction. 
Before the projective measurement on $B$, the baby universe $C$ lies outside the entanglement wedge of the individual boundary systems $A$ and $B$. 
After the measurement, in the one-sided geometry, the baby universe $C$ now lies within the entanglement wedge of $A$. 
From an information-theoretic perspective, this can be understood because the measurement outcome on $B$ is communicated to $A$ via classical communication.
In other words, through one-way LOCC (measurement on $B$ followed by classical communication from $B$ to $A$), the boundary system $A$ gains, in principle, the ability of decode the baby-universe degrees of freedom $C$. 
This suggests that operators acting on $C$ should admit a representation on $A$. 

At the same time, this seems to contradict our proposal that $C$ remains effectively independent from the viewpoint of a single boundary. A possible resolution is that, although reconstruction exists in principle, it requires decoding operations of extremely high circuit complexity for typical post-measurement Haar-random states. In this sense, $C$ may be reconstructable in theory yet operationally inaccessible in practice. Understanding this decoding complexity more precisely remains an interesting open problem.



\section{Outlook}\label{sec:outlook}

To conclude, our discussion suggests a physical mechanism by which a semiclassical baby universe can emerge with an effective entropy $S_{\mathrm{BU}}$. 
The key insight was the recent development in quantum information theory concerning the complete breakdown of complementary recovery in Haar random encoding.
This provides a potential physical mechanism for an effectively closed baby-universe region that is inaccessible to any individual boundary. 
In this interpretation, complementarity arises not as a postulate, but as a consequence of the random encoding structure. 
This perspective hints that a semiclassical baby universe need not contradict no-cloning or monogamy constraints, if the bulk reconstruction is restricted by such emergent complementarity. 


It will be interesting to investigate whether the mechanisms discussed here apply universally, for instance, to baby universes that may arise near the end of black hole evaporation or single closed universes like de Sitter spacetime.
We emphasize that our interpretation is based on information-theoretic analysis of boundary CFTs. 
How this viewpoint connects to bulk gravitational path integrals remains to be fully understood. 

\subsubsection*{von Neumann algebra and complementary recovery}

In the modern algebraic formulation of bulk reconstruction, the algebra of observables associated with a bulk region is represented as a von Neumann algebra~\cite{Leutheusser:2022bgi, Witten:2021unn}. 
A key structural property of von Neumann algebra for quantum field theories is captured by the double commutant theorem.
A closely related property is Haag duality, which states that the commutant of the algebra of observables localized in a given spacetime region coincides with the algebra associated with its complementary region. 

When complementary recovery holds, the bulk complementary algebras $a$ and $b$ are faithfully reconstructed as the corresponding boundary algebras $A$ and $B$. 
In this case, bulk Haag duality ($a$ vs. $b$) is preserved under the holographic map and translated into boundary Haag duality ($A$ vs. $B$).
Consequently, a type-III (or type-II$_{\infty}$ after including area operators) bulk algebra is mapped to a boundary CFT algebra of the same von Neumann type. 

By contrast, a breakdown of complementary recovery, as suggested by random encoding and by the semiclassical baby-universe geometry, indicates that the usual von Neumann-algebraic framework may be insufficient in such settings. 
In particular, bulk Haag duality fails to be preserved under holographic map.  
This aligns with a recent complementary perspective proposed by Liu~\cite{Liu:2025cml}, which also argues that baby-universe physics may require an extension beyond the standard von Neumann–algebraic description.
This also enables us to evade the recent no-go argument for a baby universe~\cite{Gesteau:2025obm}. 

\subsubsection*{Observer-dependence in baby universe}

Several previous works have proposed introducing observer degrees of freedom into gravitational path integral calculations in order to recover an effective baby-universe entropy $S_{\mathrm{BU}}$, even though the underlying baby-universe Hilbert space appears trivial~\cite{Abdalla:2025gzn, Harlow:2025pvj, Akers:2025ahe}. 
While such constructions indeed generate entropy, it remains conceptually unclear what these observer systems represent in the boundary CFT, or how they might emerge naturally from boundary dynamics. 

In our interpretation, observer degrees of freedom arise naturally once we include an external reference system $C$ and consider ensemble-like insertions of heavy operators.  
In quantum information theory, introducing a reference system is a standard way to study the effect of introducing measurement probes. 
In the present setting, the extended pure state $|\Psi_{ABC}\rangle$ naturally contains the label system $C$, which keeps track of which heavy operator was inserted, while the corresponding bulk baby-universe state $|\psi_c^{(i)}\rangle$ can be thought of as the observer's imprint left within the baby universe. 

From this perspective, the baby universe becomes visible not because an observer is artificially inserted, but because the heavy operator that prepares the geometry inherently serves as an observer. 
The same heavy operator that prepares the baby universe geometry also provides it with the capacity to probe the baby universe.
In this sense, it may be more appropriate to say that the baby universe is measured by an observer who created it, and thus is \emph{observer-dependent}.


It is useful to recall that the effect of introducing observers degrees of freedom on black holes in AdS/CFT has been explored from the boundary CFT viewpoint in~\cite{Yoshida:2019qqw, Yoshida:2019kyp}.
There, it has been shown that the entanglement structure of a black hole can be significantly altered by the scrambling dynamics of holographic CFTs, making some aspects of the black hole interior effectively observer-dependent. 
In particular, Hawking partner operators were reconstructed in a state-independent yet observer-dependent manner.
While it was conjectured that such effects may arise from shockwave-like backreaction near the horizon, a concrete bulk interpretation, such as one based on the gravitational path integral, remains to be understood.

Here we find a parallel between observer-dependent black hole interiors and baby universe microstates. 
In both cases, the observer is not introduced as an ad hoc bulk ingredient. 
Rather, observer-like degrees of freedom originate on the asymptotic boundaries (as operator insertions with reference systems) and influence the entanglement structure of the boundary CFT wavefunction. 
This changes entanglement structure significantly, and then manifests as an observer-dependent notion of interior, whether it be a black hole interior or a closed baby-universe region.

More recent works (e.g~\cite{Chandrasekaran:2022cip}) have also introduced observer degrees of freedom to construct type-II von Neumann algebras, with the resulting algebra depending on the observer’s worldline. 
By contrast, the notion of observer-dependence we emphasize here is stronger: the boundary manifestation of bulk operator algebras depends not only on the observer’s trajectory, but also on its microscopic quantum state, due to observer-induced backreaction. 
In our setting, observer degrees of freedom are introduced explicitly at the asymptotic boundaries and then propagated into the region of interest, inevitably inducing non-negligible backreaction. 
It is this backreaction that leads to an observer-dependent description of the interior or, in the present context, of the baby universe. 
~\footnote{
The construction of type-II von Neumann algebras provides a formal framework for describing particular aspects of semiclassical bulk gravity. However, its scope is naturally limited to the regimes where semiclassical bulk descriptions remain stable under perturbations. 
By contrast, the questions involving black hole interiors with infalling observers, or baby-universe microstates from heavy operators, require addressing observer-induced backreaction, where the underlying geometric and entanglement structures can change dramatically. 
}

\subsubsection*{Emergent non-isometricity}

The non-isometric code proposal~\cite{Akers:2022qdl} is motivated by the observation that the total bulk entropy of a black hole interior appears to exceed the Bekenstein-Hawking area at late times.
In such situations, the proposal postulates that the holographic map is effectively obtained by applying a random projection onto a subspace of dimension $e^{S_{\text{BH}}}$.
Although this picture provides a plausible resolution of the entropy–area tension, several conceptual issues remain.
What precisely enforces such a projection?
Is it a genuine dynamical process, or merely a computational device introduced to reproduce semiclassical expectations?
These ambiguities persist when analogous non-isometric treatments are invoked in the context of baby universes.

In our interpretation of the baby-universe puzzle, non-isometric encoding emerges naturally, without introducing any ad hoc projection.
When the number of heavy operators satisfies $M \gg e^{S_{\mathrm{BU}}}$, the baby-universe microstates ${|\psi^{(i)}_c\rangle}$ become overcomplete relative to the total Hilbert-space dimension $e^{S_{\mathrm{BU}}}$.
In previous approaches, one explicitly projected these states onto an $e^{S_{\mathrm{BU}}}$-dimensional subspace by hand~\cite{Harlow:2025pvj, Akers:2025ahe}.
Here, the same effective projection arises automatically due to the Euclidean (imaginary-time) evolution contained in $\tilde{O}^{(i)} = e^{-\frac{\beta H}{2}} O^{(i)} e^{-\frac{\beta H}{2}}$.
Thus, non-isometricity is built into the microscopic construction of the state itself.

In fact, one may generalize the AS$^{2}$ setup to construct a two-sided black hole with large interior entropy by evolving the system in Euclidean time  at a low, but still above the Hawking-Page temperature.
Choosing $M \gg e^{S_{\text{BH}}}$ produces a situation that would seemingly violate the area law.
However, there is no paradox as the Euclidean evolution automatically projects the dynamics onto an $e^{S_{\text{BH}}}$-dimensional subspace.

In an actual evaporating black hole, apparent non-isometricity arises without any explicit Euclidean evolution. 
Instead, unitary evaporation itself must dynamically realize a quantum-gravitational process that appears as a projection at the semiclassical level. Still, the toy extension described above may provide a tractable model in which the origin of non-isometricity is clear.

\subsection*{Acknowledgment}

We thank 
Ahmed Abdalla,
Chris Akers,
Stefano Antonini,
Luca Ciambelli,
Kenny Higginbotham,
Zhi Li,
Pratick Rath,
Brian Swingle,
Tomonori Ugajin,
Zhencheng Wang,
and
Zixia Wei
for useful discussions.
Research at Perimeter Institute is supported in part by the Government of Canada through the Department of Innovation, Science and Economic Development and by the Province of Ontario through the Ministry of Colleges and Universities. 
This work was supported by JSPS KAKENHI Grant Number 23KJ1154, 24K17047.

\appendix

\section{Tripartite random entanglement}\label{sec:theorem}

Here we quote technical results from~\cite{Li:2025nxv}.

\subsection{Local unitaries}

Consider a tripartite Haar random state $|\Psi_{ABC}\rangle$.
For a non-negative constant $0< h \leq 1$, the one-shot LU-distillable entanglement is defined as
\begin{align}\label{def-EDLU}
\ED^{\text{[LU]}}_{h}(A:B) 
\equiv \sup_{m \in \mathbb{N} }\sup_{\Lambda\in \text{LU}} \Big\{ 
 m  \Big| 
\Tr\big( \Lambda(\rho_{AB}) \Pi^{[\text{EPR}]}_{R_{A} R_B} \big) \geq h^2 
  \Big\},
\end{align}
where $\Lambda = U_A \otimes U_B \in \text{LU}$ represents a local unitary acting on $A\otimes B$, and $\Pi^{[\text{EPR}]}_{R_{A} R_B}$ is a projection operator onto $m$ EPR pairs supported on $R_A,R_B$. 
Here, $R_A \subseteq A$ and $R_B\subseteq B$, and $|R_A|=|R_B|=m$ with $|R_A|,|R_B|$ denoting the number of qubits in subsystems $R_A$ and $R_B$ respectively. 
The parameter $h$ controls the fidelity of EPR pairs, where $h =1$ corresponds to perfect EPR pairs while $h \approx 0$ corresponds to low fidelity EPR pairs.

\begin{theorem}\label{theorem:mainLU}
    If $\delta\defeq h^2-2^{-2m}>0$, then for an arbitrary constant $0<c<2$, we have
    \begin{equation}\label{eq:bound}
        \log\prob{\ED^{\text{\emph{[LU]}}}_{h}(A:B) \geq m}
        \leq -c\delta^2d+O\qty((d_A^2+d_B^2)\log \frac{1}{\delta}).
    \end{equation}
\end{theorem}


Theorem~\ref{theorem:mainLU} provides a meaningful bound when the second term of  is subleading, ensuring that the right-hand side is negative.
The bound then implies that EPR pairs cannot be LU distilled from $\rho_{AB}$ with a fidelity better than that from a maximally mixed state.
In other words, any attempt to enhance the EPR fidelity by applying LU transformations $U_A \otimes U_B$ will be useless! 
A similar bound can be obtained for LO-distillable entanglement, see~\cite{Li:2025nxv}. 



\subsection{Logical operators}

Given two parameters $0\leq h,w\leq 1$, we say an isometric encoding $V : C \to AB$ admits a logical unitary operator $U_{AB}$ if there exists a unitary $U_C$ such that
\begin{equation}\label{eq:logicalrequire}
    \left| \frac{\Tr(U_C^\dagger V^\dagger U_{AB} V)}{d_C} \right|\geq h^2 \text{~~and~~} \frac{|\Tr(U_C)|}{d_C}\leq w.
\end{equation}

The first inequality imposes a fidelity $h$ on a logical unitary operator $U_{AB}$.
Namely, it compares two isometries $U_{AB} V$ and $VU_C$ via the fidelity between their Choi states:
\begin{equation}
    \frac{\Tr(U_C^\dagger V^\dagger U_{AB} V)}{d_C}
    =\bra{EPR} U_C^\dagger V^\dagger U_{AB} V \ket{EPR}
    =\bra{\Psi}U_{AB}\otimes U_C^*\ket{\Psi},
\end{equation}
where $\ket{\Psi}=(V\otimes I)\ket{EPR}$ is the Choi state for $V$.
Note that
\begin{equation}
    1-|\bra{EPR} U_C^\dagger V^\dagger U_{AB} V \ket{EPR}|\leq \frac{1}{2}\norm{U_{AB} V-VU_{C} }_\infty^2,
\end{equation}
thus, a large $h$ is also a necessary condition for $U_{AB} V$ and $VU_C$ to be close in the operator norm.

The second inequality ensures the non-triviality of the logical operator.  
Namely, $w<1$ is required to ensure that $U_C$ acts non-trivially on the input state.
Otherwise, $w=1$ would imply that $U_C$ is a phase $e^{i\theta}I_C$, and $e^{i\theta}I_{AB}$ is a trivial logical operator for $e^{i\theta}I_C$ with $h=1$.
On the other hand, if $U_C$ is a unitary conjugation of a Pauli operator, then $w=0$.

\begin{theorem}\label{thm:LOGICAL}
    Assuming $\delta\defeq h^2-w>0$, there exists an absolute constant $c^\prime>0$, such that for a Haar random isometry $V$,
    \begin{equation}
        \log\prob{V\text{~\emph{admits a logical operator on}~} A}
        \leq -c^\prime\delta^2 d+O\left((d_A^2+d_C^2)\log \frac{1}{\delta}\right).
    \end{equation}
\end{theorem}


\providecommand{\href}[2]{#2}\begingroup\raggedright\endgroup

%
\end{document}